# Low Microwave Surface Resistance in $NdBa_2Cu_3O_{7-\delta}$ Films Grown by Molecular Beam Epitaxy

Jose KURIAN* and Michio NAITO

*NTT Basic Research Laboratories, NTT Corporation, 3-1 Morinosato Wakamiya, Atsugi-shi, Kanagawa 243 0198, Japan*

Abstract

We report the growth of $NdBa_2Cu_3O_{7-\delta}$ films on (100) MgO substrate by Molecular Beam Epitaxy (MBE). Large area $NdBa_2Cu_3O_{7-\delta}$ films with homogeneous superconducting properties were grown by precise control of stoichiometry and the optimisation of growth parameters. The stoichiometric ratio of Nd:Ba:Cu close to 1:2:3 yields films with $T_C$ of 94 K and $J_C$ values above 3.5 MA/cm$^2$ at 77 K on bare MgO substrate. The $NdBa_2Cu_3O_{7-\delta}$ films grown under optimised conditions had excellent in-plane texture and good metallicity. The most significant characteristic of our MBE grown $NdBa_2Cu_3O_{7-\delta}$ films is the very low microwave surface resistance values at all temperature range compared to its $YBa_2Cu_3O_{7-\delta}$ counterpart with typical value of ~870 μΩ at 77 K & 22 GHz. Our results on the MBE grown $NdBa_2Cu_3O_{7-\delta}$ films suggests that $NdBa_2Cu_3O_{7-\delta}$ is a superior choice for the realisation of commercial microwave applications.

---

* E-mail address: jkurian@will.brl.ntt.co.jp

The growth of high quality films of high temperature superconductors (HTSC) is of significance from both technological as well as basic science point of view. The 123 family of superconductor films have shown tremendous potential for application in microwave devices and circuits. Due to the low microwave surface resistance ($R_S$) of 123 films, microwave devices such as resonators, filters, delay lines, etc., with high Q values and low loss can be fabricated. Among the different copper oxide superconductors discovered so far, the $YBa_2Cu_3O_{7-\delta}$ (YBCO) superconductor has been the material of choice for application as the deposition and patterning processes are extensively studied and established, and films with $T_C$ of ~90 K can be routinely fabricated by different deposition techniques. An alternative material to YBCO is the $NdBa_2Cu_3O_{7-\delta}$ (NBCO) superconductor, which has certain advantages over YBCO superconductor. NBCO has distinctly higher $T_C$ than YBCO and has better stability, better crystallinity and superior surface characteristics.[1-6] Recent studies reveal the high stability of NBCO against water corrosion compared to other 123 superconductors.[6] Presently different deposition techniques have been employed for the growth of 123 superconductor films such as pulsed laser deposition,[7-9] sputtering,[10] molecular beam epitaxy (MBE),[11] metalorganic chemical vapour deposition,[12] etc. Among the different deposition techniques described above, MBE has the specific advantage that large area films with excellent homogeneous properties over the entire wafer can be prepared and at relatively low substrate temperatures. The low substrate temperature used in the MBE growth becomes significant in the case of growth of 123 films on substrates such as MgO, since higher substrate temperature leads to interfacial reaction between the substrate and the film resulting in the degradation of superconducting properties. Even though there have been many studies regarding the surface characteristics and stability of NBCO which are distinctly different from other 123 superconductors, the microwave characteristics of the NBCO films have not been investigated in detail. In this letter we report the growth of NBCO films on (100) MgO by MBE with high $T_C$ of 94 K and $J_C$ above 3.5 MA/cm$^2$ at 77 K. The most significant characteristics of our NBCO films grown by MBE is the distinctly low microwave surface resistance values at all temperature range than the YBCO counter part with typical value of ~870 μΩ at 77 K and 22 GHz. The $R_S$ values of MBE grown NBCO films were at least a factor of two lower than that of best commercially available YBCO films and is the lowest $R_S$ value reported in literature for 123 films. The combination of higher $T_C$, higher $J_C$ and low $R_S$ of NBCO makes it a superior choice for microwave applications.

NBCO thin films were grown on (100) MgO substrates by MBE in a custom designed ultra high vacuum evaporation chamber (base pressure ~10$^{-8}$ Torr) with load lock chamber. The MBE chamber used in the present study was equipped with RF activated oxygen source for *in situ* oxygenation and the real time monitoring of the growth process can be performed by reflection high energy electron diffraction (RHEED). The thin film growth was carried out using e-beam evaporation from Nd, Ba & Cu metal sources. The precise control of the individual rates of the constituent elements was achieved by a feedback system using electron impact emission spectrometric technique. More details of our growth set-up can be found in references 13 and 14. The MgO substrates used were pre-annealed in oxygen atmosphere at 1000 $^o$C for 10 h, as it is reported that the pre-annealing step improves the surface quality of MgO substrates.[15-17] The MgO substrates were heated radiantly and the substrate temperature was monitored by a thermocouple and calibrated by an optical pyrometer. NBCO films were grown under RF activated oxygen with a flow rate of 2 sccm and an RF power of 300 W and the resulting run pressure was ~6 x $10^{-5}$ Torr. The growth rate was ~0.4 nm/s. After the termination of the growth process, the films were cooled down to ~150 $^{o}$C in radical oxygen under the same flow rate. In order to full loading of oxygen, the films were annealed in flowing oxygen for 30 min at 320 $^{o}$C in an external furnace. In the present set-up, we can grow double-sided films with diameter up to 4″.

In order to grow high quality superconducting NBCO thin films, the most vital parameter is the precise rate control so as to grow NBCO films with correct stoichiometry.[18, 19] Figure 1(a) shows a typical example of the effect of Cu:Nd ratio on the different properties of NBCO films (Ba:Nd ratio ~2, substrate temperature 700 $^{o}$C & films thickness 250 nm) highlighting the importance of composition. The actual composition of the films was determined by inductively coupled plasma (ICP) analysis. NBCO films grown with Cu:Nd ratio slightly lower than 3 ('Cu-poor' case) resulted in the drastic reduction of $T_C$. Also 'Cu-poor' films showed poor crystallinity as reflected in the low XRD intensity of NBCO peaks. As the Cu:Nd ratio reaches 3, the $T_C$ of NBCO films peaks and a further increase in Cu:Nd ratio up to ~3.6 shows a saturation behaviour in $T_C$ and $\rho_{RT}$. The crystallinity of NBCO films also shows a saturation trend on the 'Cu-rich' side (Cu:Nd ratio > 3). Substantially excess Cu (Cu:Nd > 3.6) give rise to a drop in $T_C$ and increase in $\rho_{RT}$. The studies carried out by varying the Ba:Nd ratio show that the highest $T_C$ with low resistivity is observed at Ba:Nd ratio of ~2. Figure 1(b) shows the effect of Ba:Nd ratio on $T_C$ and room temperature resistivities of NBCO films (substrate temperature 700 $^{o}$C & film thickness 250 nm). One of the distinct point worth noting is that the 'Ba-rich' films (Ba:Nd > 2) shows low room temperature resistivity ($\rho_{RT}$), but the $T_C$ decreases as the Ba:Nd ratio increases from 2. Our studies indicate that the NBCO films grown with composition close to 123 gives highest $T_C$ with sharp transition, excellent metallicity ($\rho_{300K}/\rho_{100K}$ ~3) and low resistivity. The effect of substrate temperature on $T_C$ and $\rho_{RT}$ of NBCO films are shown on Fig. 1(c). It can be seen from Fig. 1(c) that the $T_C$ of NBCO films reaches a maximum at ~690 $^o$C and a further increase in growth temperature does not show any effect on $T_C$ or $\rho_{RT}$.

The characteristics of the optimised NBCO films grown on MgO are presented in Figs. 2 and 3. The structure of the NBCO films was examined by X-ray diffraction (XRD) using a Rigaku X-ray diffractometer (RINT, Japan) with Cu-K$_\alpha$ radiation and Fig. 2(a) shows the typical 2θ-θ XRD pattern of 500 nm thick NBCO thin film grown on (100) MgO substrate. In the XRD pattern, all the peaks except the characteristic peak of MgO are that of (00*l*) reflection of NBCO. The crystallinity of the NBCO films was examined by employing ω-scan of (005) reflection of NBCO and the full width at half maximum of the rocking curve of (005) reflection of NBCO films was 0.24 degrees indicating the high crystalline quality. The excellent in-plane orientation of the NBCO films on MgO is evident from the ϕ-scan measurement of (103) reflection of NBCO (inset of Fig. 2(a)). The RHEED and SEM images of a 500 nm thick NBCO films is shown in Fig. 2(b) and Fig. 2(c) respectively, indicating the smooth surface morphology of the NBCO films. The smooth surface morphology is one of the distinct characteristics of NBCO films compared to other 123 superconductor films. The RHEED set-up in our MBE growth chamber enables the real time monitoring of the growth process, which is helpful in identifying any deviations from the required stoichiometry and making real time corrections.[20]

The superconductivity of the NBCO films was studied by standard four-probe technique. Figure 3 shows the typical variation of resistivity with temperature for a 500 nm thick NBCO film on (100) MgO. The NBCO films with stoichiometry close to 123 gives a room temperature resistivity between 140 to 180 $\mu\Omega$.cm. Such NBCO films shows good metallic behaviour in the normal state ($\rho_{300K}/\rho_{100K}$ ~3) and gave a $T_C(0)$ of 94 K with a transition width below 1 K. The critical current density $J_C$ of the NBCO films grown on (100) MgO substrate was determined by induction method which shows that the NBCO films have $J_C$ values above 3.5 MA/cm$^2$ at 77 K with high degree of homogeneity. One of the significant point to be noted is the high degree of reproducibility of these results on bare MgO substrates which may be attributed to the MBE growth process.

The microwave surface resistance of NBCO films was measured by dielectric resonator method with sapphire rod and an impedance analyser (Agilent 8510C). A pair of 20 mm × 20 mm size NBCO films grown on (100) MgO was used for the $R_S$ measurements. Our studies on the microwave $R_S$ of NBCO films show that NBCO films have distinctly low $R_S$ values throughout the studied temperature region than its YBCO counterpart. Figure 4. shows the temperature dependence of microwave $R_S$ of MBE grown NBCO films on MgO substrate measured at 22 GHz. It was found that our MBE grown NBCO films on MgO gives $R_S$ values of 800 - 1000 $\mu\Omega$ at 77 K and ~ 200 $\mu\Omega$ at 20 K. The common $R_S$ values reported in literature for YBCO films are ~1500 $\mu\Omega$ and ~250 $\mu\Omega$ at 77 K and 20 K respectively (scaled to 22 GHz).[21,22] The $R_S$ values our NBCO films at 20 K is at least a factor of 2 lower than that of best commercially available films which was measured in our measurement set up under identical conditions. Some of the NBCO films show $R_S$ values of 100 $\mu\Omega$ at 20 K even though they show a slightly higher $R_S$ (1000 - 1200 $\mu\Omega$) at 77 K. Studies are underway to identify the cause. We speculate that this may be due to slight variations in composition.

In conclusion, we report the growth of high quality NBCO films on MgO by MBE. High quality NBCO films could be grown by the precise control of stoichiometry and optimisation of growth parameters. NBCO films grown under the optimised conditions gave a $T_C$ (0) of 94 K and a $J_C$ above 3.5 MA/cm$^2$ at 77 K with room temperature resistivity as low as 140 $\mu\Omega$.cm and good metallicity. Furthermore, they have smooth surface characteristics. The most significant characteristic of our MBE grown NBCO films is the low $R_S$ values when compared to its YBCO counterpart. The higher $T_C$, $J_C$ and the low $R_S$ of the NBCO films makes it a superior choice for commercial microwave applications.

**Acknowledgements**

The authors like to acknowledge Prof. Y. Kobayashi and T. Hashimoto of Saitama University and H. Obara and M. Murugesan of AIST, Japan for fruitful discussions. The authors also like to acknowledge H. Sato, H. Yamamoto, S. Karimoto, K. Ueda and A. Tsukada of NTT Basic Research Laboratories for their valuable suggestions and discussions at various stages of this work.

**Figure captions**.

Fig. 1. (a) Variation of $T_C$ (onset) & $T_C$ (0) and room temperature resistivity of NBCO films with (a) Cu:Nd ratio, (b) Ba:Nd ratio and (c) substrate temperature. (The lines provided are guide for the eye.)

Fig. 2. (a) Typical 2θ-θ XRD pattern of NBCO film grown on (100) MgO by MBE (in-set shows the φ-scan XRD pattern (103) reflection of NBCO film), (b) RHEED pattern in (100) direction of NBCO and (c) the SEM image of 500 nm thick NBCO film on MgO.

Fig. 3. Typical temperature-resistivity curve of NBCO film grown on MgO and the in-set shows the enlarged portion of the superconducting transition.

Fig. 4. Variation of microwave $R_S$ with temperature of a 500 nm thick NBCO film grown on MgO by MBE at 22 GHz.

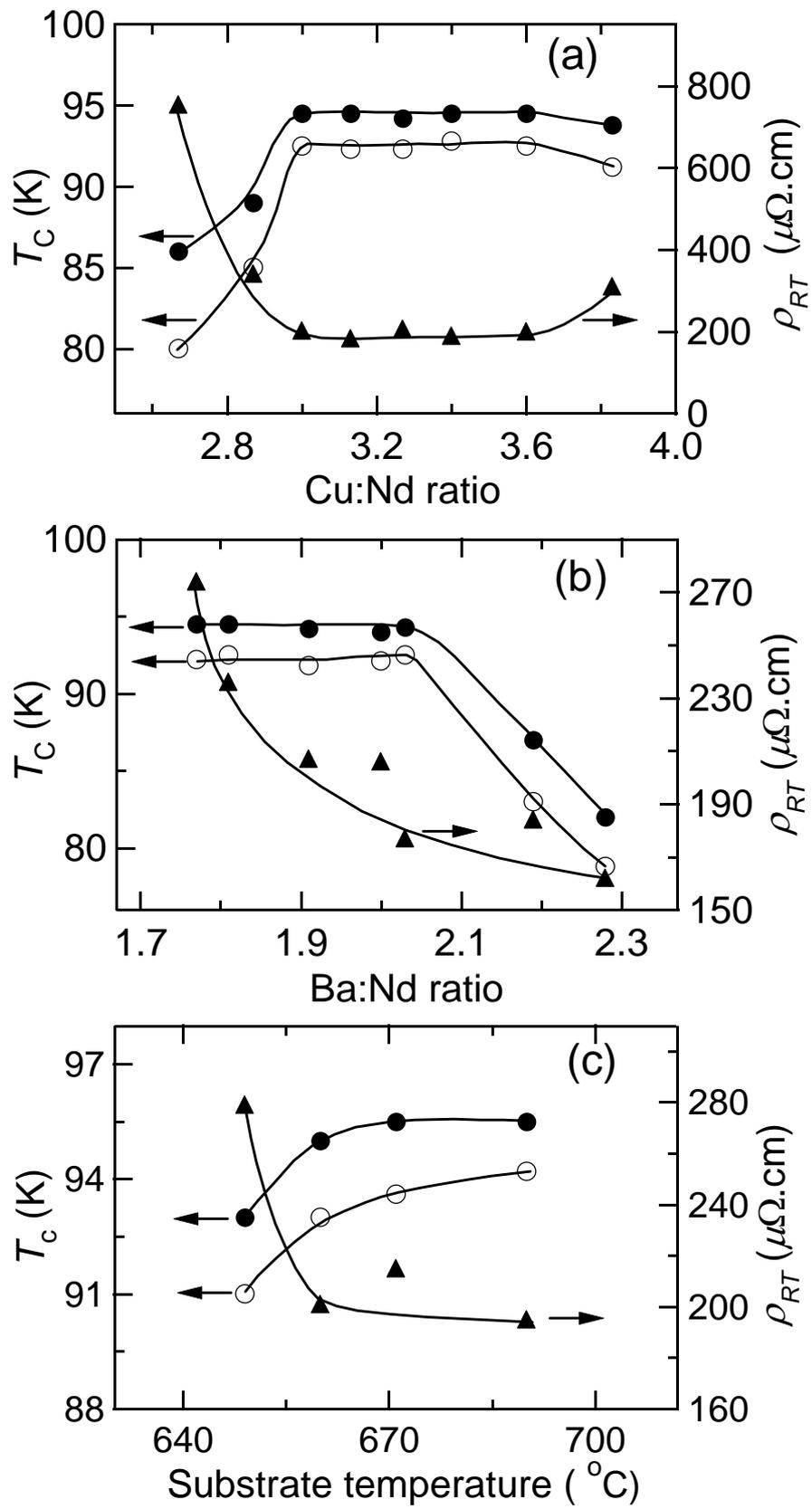

Fig.1

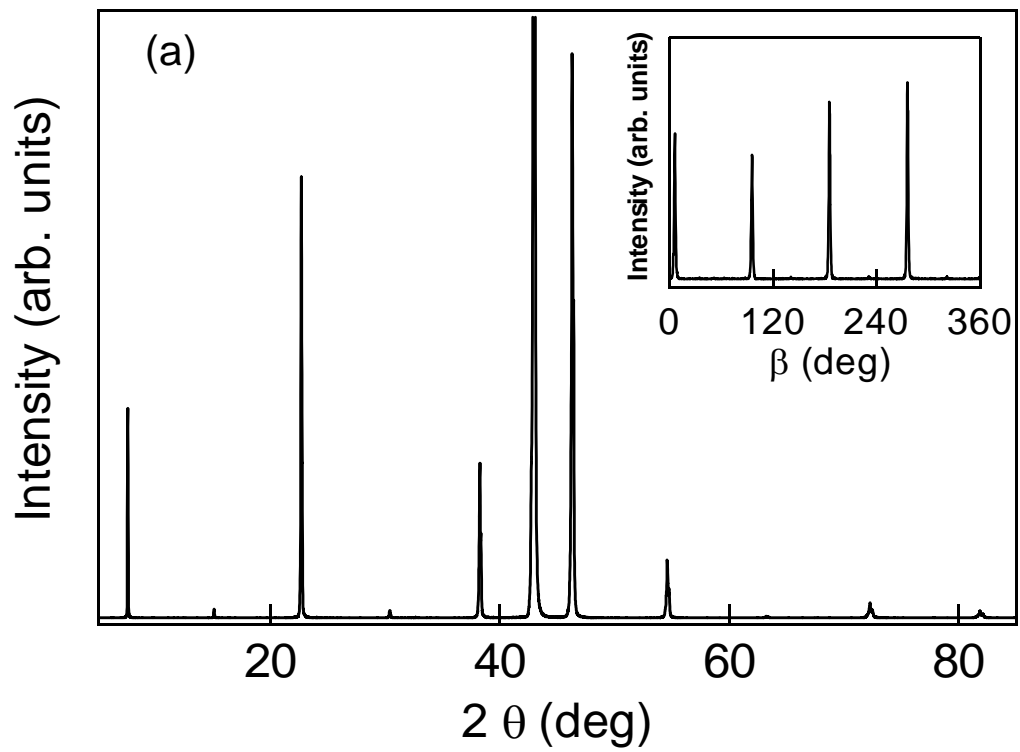

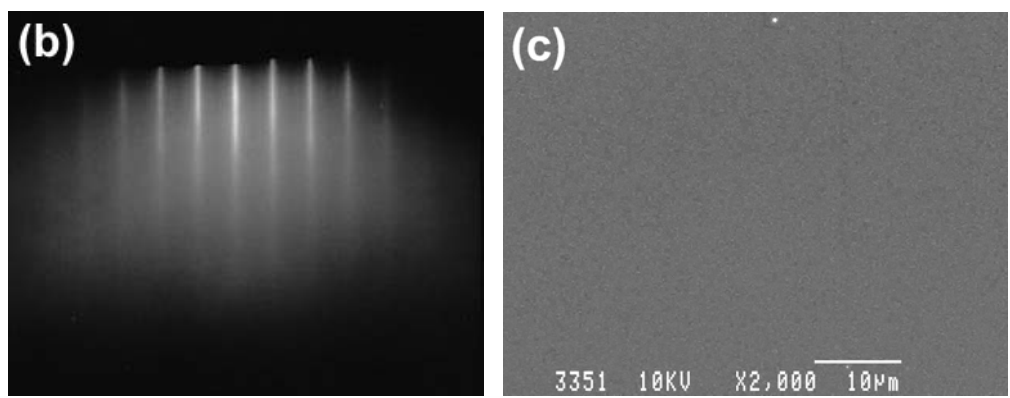

Fig. 2

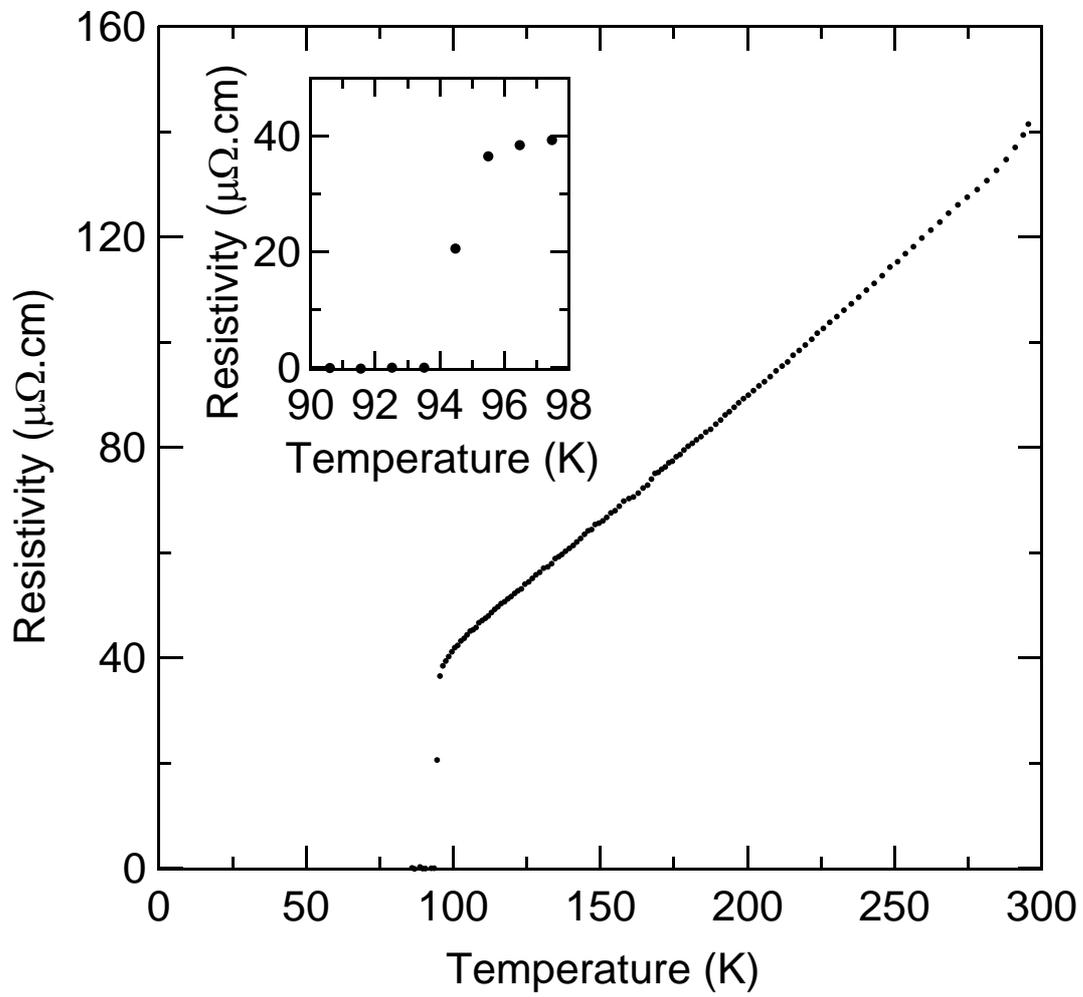

Fig. 3

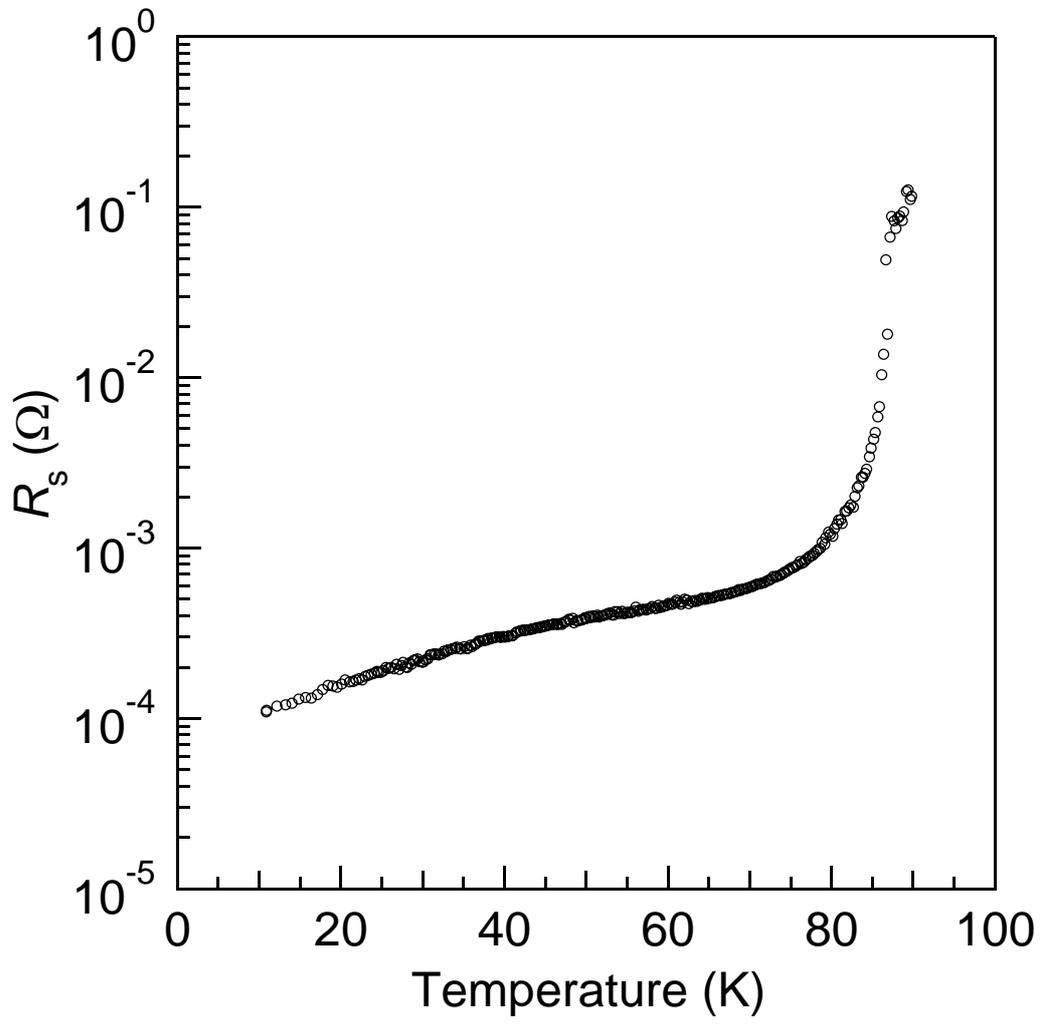

Fig. 4